\documentclass[12pt]{article}

\usepackage{amsmath,amssymb,graphicx,color}

\newwrite\ffile\global\newcount\figno \global\figno=1

\def\writedef#1{}

\input epsf
\def\figin{\epsfcheck\figin}\def\figins{\epsfcheck\figins}
\def\epsfcheck{\ifx\epsfbox\UnDeFiNeD
\message{(NO epsf.tex, FIGURES WILL BE IGNORED)}
\gdef\figin##1{\vskip2in}\gdef\figins##1{\hskip.5in}% blank space instead
\else\message{(FIGURES WILL BE INCLUDED)}%
\gdef\figin##1{##1}\gdef\figins##1{##1}\fi}

\def\figinsert{}
\def\ifig#1#2#3{\xdef#1{fig.~\the\figno}
\writedef{#1\leftbracket fig.\noexpand~\the\figno}%
\figinsert\figin{\centerline{#3}}\medskip\centerline{\vbox{\baselineskip12pt
\advance\hsize by -1truein\center\footnotesize{  Fig.~\the\figno.}
#2}}
\bigskip\endinsert\global\advance\figno by1}
\def\endinsert{}
\providecommand{\Tr}{}
\renewcommand{\Tr}{\ensuremath{\mathrm{Tr}}}

\usepackage{amssymb}
\usepackage{graphics}

\begin{document}

%\begin{flushright}
%ITP-Hep-02/06 \\
%hep-th/000000 \\
%March 2006 \\
%\end{flushright}
\begin{center}
\vskip 2cm \Large {\bf Three Flavour QCD from the Holographic
Principle} \vskip 1.5cm \centerline{ \normalsize \bf Jonathan P.
Shock \footnote[1]{\noindent \tt
 jps@itp.ac.cn} and Feng Wu \footnote[2]{\noindent \tt
 fengwu@itp.ac.cn}}
\vskip 1cm \small
{\em Institute of Theoretical Physics\\
 Chinese Academy of Sciences\\
 P.O. Box 2735\\
 Beijing 100080, CHINA}
\end{center} \vskip 2cm
\begin{abstract}
\noindent Building on recent research into five-dimensional
holographic models of QCD, we extend this work by including the
strange quark with an $SU(3)_L\times SU(3)_R$ gauge symmetry in
the five-dimensional theory. In addition we deform the naive $AdS$
metric with a single parameter, thereby breaking the conformal
symmetry at low energies. The vector and axial vector sectors are
studied in detail and both the masses and decay constants are
calculated with the additional parameters. It is shown that with a
single extra degree of freedom, exceptional agreement with
experimental results can be obtained in the light quark sector
while the kaon sector of the vector and axial vector octet is
found to give around $10\%$ agreement with lattice results. We
propose some simple extensions to this work to be taken up in
future research.
\end{abstract}
\thispagestyle{empty}

\newpage
\pagenumbering{arabic}

\section{Introduction}
Since 1973  QCD \cite{Phys.Lett.B47.365} has been the established
theory of the strong interactions. Though understood in the
perturbative regime, there remains no analytic solution for the
phenomenon of confinement for a general Yang-Mills theory.
Effective chiral perturbation theory is constructed in order to
describe the low energy interactions of matter fields. A
systematic method to handle the non-perturbative regime of the
strong interaction from first principles has been sought though
without success so far. Inspired by the gauge/gravity duality
\cite{hep-th/9711200,hep-th/9802109,hep-th/9802150} and the
success of flavoured models from ten-dimensional supergravity
\cite{hep-th/0205236,hep-th/0304032,hep-th/0306018,hep-th/0311270,hep-th/0403279},
several holographic models
\cite{hep-ph/0501128,hep-ph/0501218,hep-ph/0507049,hep-ph/0510240,hep-ph/0510268,hep-ph/0510334,hep-ph/0512089,hep-ph/0512240,hep-ph/0612077,
hep-ph/0603249} have been proposed recently with surprising
success. The key point is that although QCD is not a conformal
field theory and its string dual is still unknown, it approaches
the conformal limit in the ultraviolet regime. Low energy
properties such as confinement and chiral symmetry breaking are
introduced in these models in a simple manner. The models
discussed so far have all included quarks in $SU(2)$ or $SU(3)$
flavour groups, each with an identical mass.

Since the mass differences between the mesons in the same
octet are small compared with the scale $\Lambda_{QCD}$ for the
lowest vector and axial-vector mesons, it is interesting to study
whether the same successful results of the isovector sector can be
achieved after including the third flavour. We will show that
indeed this is the case. In this letter we will extend the work of
\cite{hep-ph/0501128} to three flavours without assuming the
flavour symmetry and calculate the mass spectrum and decay
constants of $1{}^3 S_1$ vector mesons and $1 {}^3 P_1$
axial-vector mesons. We will not discuss the scalar sector and the
$U(1)_A$ problem but will leave these questions for future
research.

The metric in the five-dimensional bulk with four-dimensional
Poincar$\acute{e}$ symmetry and a compact extra dimension is
\begin{equation}
ds^2=\theta (z) \theta(z_{IR}-z) a^2(z)(dx^{\mu}dx_{\mu}-dz^2),
\end{equation}
where
\begin{equation}
a^2(z)=\frac{1}{z^2}(1+\sum_{i=1}^n \alpha_n z^{2n}) \hspace{1cm} \mbox{and $\theta(z)$ is the step function}.
\end{equation}
The $5^{th}$ dimension corresponds to the energy scale in the
four-dimensional theory. The metric has a double pole on the UV
boundary ($z=0$) and is asymptotically $AdS_5$ in the
$z\rightarrow 0$ limit. The IR boundary $Z_{IR} \sim
O(\Lambda_{QCD}^{-1})$ is put in by hand in this model in order to
introduce the mass gap and can be interpreted as the position of
an IR D-brane. In this work we will consider both the simplest
case, that is, the $AdS$ bulk such that $a(z)={1\over z}$ and the
case where only $\alpha_{1}$ is non-zero in the warp factor
$a(z)$. Since the $AdS$ background is dual to $\mathcal{N}=4$ SYM,
considering the deformed metric is natural for a model builder as
QCD is not a conformal field theory. In the bulk the gauge group
is chosen to be $SU(3)_L \otimes SU(3)_R$, which is dual to the
global chiral group on the UV boundary where QCD behaves
approximately like a CFT. In addition to the gauge fields
$L_{M}(z,x)$ and $R_{M}(z,x)$, a scalar field $X$ transforming as
$(3_L,3_R)$ is also introduced to induce the chiral symmetry
breaking. The non-renormalizable five-dimensional action is
\begin{equation}
S=\int d^5x
\sqrt{g}\Tr\left\{-\frac{1}{4g_5^2}\left(L_{MN}L^{MN}+R_{MN}R^{MN}\right)+|DX|^2+3|X|^2\right\},
\end{equation}
where
\begin{eqnarray}
D_{M} X &=& \partial_{M} X -i L_{M} X+ i X R_{M},\nonumber\\
L_M&=&L_{M}^{a} t^a,\nonumber\\
L_{MN}&=& \partial_{M} L_N -\partial_N L_{M} -i [ L_M, L_N],
\hspace{1cm} \mbox{similarly for } R,
\end{eqnarray}
where $\Tr[t^a, t^b]={1\over 2} \delta^{ab}$. For our interest in
this letter, we keep only terms up to quadratic order in fields. The effects of including higher order non-renormalizable terms is equivalent to
deforming the metric from the $AdS$ metric, which will also be
considered in this work. However, the $\alpha_{n}$ terms are blind to flavours.
The only source which violates $SU(3)_{flavour}$ symmetry comes from $X$. We define the vector and axial-vector
gauge bosons to be $V_M={1\over2}(L_M+R_M)$ and $A_M={1\over2}
(L_M-R_M)$ respectively. Following \cite{hep-ph/0501128}, we will
choose the axial gauge $V_z=A_z=0$. According to the holographic
recipe \cite{hep-th/9802150, hep-th/9905104}, the non-normalizable
mode of $V_{\mu}$ is dual to the conserved vector current in QCD
while the normalizable modes give the five-dimensional extension
of the vector meson wavefunctions, similarly for $A_{\mu}$ fields.
Also, the mass of the scalar field $X$ is determined by the
AdS/CFT correspondence. First, we consider the pure $AdS$ case.
After solving the equation of motion for the expectation value of
$X$, the solution is
\begin{equation}
\langle X(z) \rangle = {1 \over 2} (M z + \Sigma z^3),
\end{equation}
where
\begin{equation}
M= \lim_{z\to 0}2{\langle X(z) \rangle  \over z}
\hspace{1cm}\mbox{and}\hspace{1cm} \Sigma= { 2\langle X(z_{IR})
\rangle -M z_{IR} \over z_{IR}^{3}}.\nonumber
\end{equation}
$M$ is the quark mass matrix and $\Sigma=\langle \bar{q} q
\rangle$, which can be easily shown by matching the
four-dimensional effective Lagrangian to the chiral Lagrangian
\cite{hep-ph/0501218}. $\Sigma$ is responsible for the $\chi SB$
in the chiral limit and we expect $\Sigma \propto \mathbf{1}$ in
this limit. That is, $\Sigma \sim \alpha \mathbf{1} + O(M)$.
Instead of considering possible boundary terms which determine the
specific forms of $M$ and $\Sigma$, we will parametrize the
effects of these terms and choose $M=\mathit{diag}(m, m, m_{s})$
in the isospin limit for phenomenological interest and
$\Sigma=\mathit{diag}(c,c,c_s)$. With this choice, $SU(3)_L\otimes
SU(3)_R$ is broken to $SU(2)\otimes U(1)$.

Substituting $X=\langle X \rangle e^{i2t^a \pi^a (x,z)}$ back into
the action, one can show that the five-dimensional fields
$V_{\mu}^{a}$ and $ A_{\mu}^{a}$ have block-diagonal z-dependent
mass matrices $M_{V}$ and $M_{A}$ respectively where
\begin{equation}
M_{V}^{2}=\left(
\begin{array}{lll}
 \mathbf{0_{3 \times 3}} &0 &0 \\
 0 & {1\over 4} \left( m-m_s +(c-c_s)z^2 \right)^2 z^2 \mathbf{1_{4\times 4}} & 0 \\
 0& 0 & 0
\end{array}
\right),
\end{equation}
and
\begin{equation}
M_{A}^{2}=\left(
\begin{array}{lll}
 (m+c z^2 )^2 z^2 \mathbf{1_{3 \times 3}} &0 &0 \\
 0 & {1\over 4} \left( m+m_s +(c+c_s)z^2 \right)^2 z^2 \mathbf{1_{4\times 4}} & 0 \\
 0& 0 & {1\over 3} \left( 2\left( m_s+c_s z^2 \right)^2 +\left( m + cz^2\right)^2 \right)^2 z^2
\end{array}
\right).
\end{equation}
$V_{\mu}^{1,2,3}$, $V_{\mu}^{4,5,6,7}$, and $V_{\mu}^{8}$
correspond to isovector, isodoublet, and isosinglet vector mesons
in the octet of the quark model, similarly for the axial-mesons.
Note that the mass of isovector and isosinglet vector mesons,
which are related to $M_{V}$, are the same and independent of
$\langle X \rangle$. In the large $N_c$ limit
\cite{Nucl.Phys.B72.461}, where Zweig's rule can be shown to be
exact, this result suggests the isosinglet in the octet to be the
$\omega$ meson in the ``ideal mixing" limit. The origin of the
mass for $V_{\mu}^{a}$ is not from the Higgs mechanism but only
from the explicitly flavour symmetry breaking term. In the
$SU(3)_{\mbox{flavour}}$ limit, $V_{\mu}^{a}$ fields are massless.
In this limit, all the vector mesons in the same octet have the
same mass.

The coupling constant $g_{5}$ is determined by matching the high
energy expansion for the two-point function, calculated by the
holographic recipe, to the same function calculated from the OPE
in large $N_c$ limit \cite{Nucl.Phys.B147.385}. The result is
\cite{hep-ph/0501128} $g_{5}^{2}={12\pi^2 \over N_{c}}$. This
result relates the coupling constant $N_c$ in large $N_c$ QCD to
the one in its dual model.

The vector meson, $V$, wavefunction, $\Phi_{V}$, and the axial
meson, $A$, wavefunction, $\Phi_{A}$, which are the normalizable
modes of four-dimensional Fourier transformed transverse field
$V_{\mu \perp}^{a}$ and $A_{\mu \perp}^{a}$ respectively, satisfy
the following linearized equations of motion (no summation)
\begin{equation}
\left[ \partial_{z}^2 +\partial_{z}\left(\ln
a(z)\right)\partial_z+\left( q^2 -( g_{5}^{2} a(z)^2
M_{V}^{2})_{aa} \right) \right]
\Phi_{V}^{a}(q,z)=0,\label{eq.phiV}
\end{equation}
and
\begin{equation}
\left[ \partial_{z}^2 +\partial_{z}\left(\ln
a(z)\right)\partial_z+\left( q^2 -( g_{5}^{2} a(z)^2
M_{A}^{2})_{aa} \right) \right]
\Phi_{A}^{a}(q,z)=0,\label{eq.phiA}
\end{equation}
with boundary conditions $\partial_{z} \Phi_{V}^{a}(q, z_{IR})=0$
and $ \Phi_{V}^{a}(q, \epsilon)=0$, similarly for $\Phi_{A}^{a}$.
In the large $N_c$ limit, the hadron masses can be obtained from
the poles of the two-point function. This is equivalent to solving
Eq.(\ref{eq.phiV}) and Eq.(\ref{eq.phiA}) with eigenvalues
$q^2=m_{V}^{2}$ and $q^2=m_{A}^{2}$ respectively. The decay
constants $F_{V,A}$, which are defined by $\langle 0 \mid
J_{{V,A}\mu}^{a} \mid V^{b}, A^{b} \rangle = \delta^{ab} F_{V,A}
\epsilon_{\mu}$ with $\epsilon_{\mu}$ the polarization vector, can
be shown to be \cite{hep-ph/0501128}
\begin{equation}
F_{V,A}^{2} = {1\over g_{5}^{2}} \left( {\Phi_{V,A}^{''} (0) \over
N} \right)^2, \label{eq.VAdecay}
\end{equation}
where
\begin{equation}
N= \int_{0}^{z_{IR}} dz a(z) \vert \Phi_{V,A}(z) \vert^2.\nonumber
\end{equation}

The masses of pseudoscalar mesons $P$ can be obtained by solving
the following equations mixed by the normalizable modes of the
fields $\pi^{a}(x,z)$ and the longitudinal components of
$A_{\mu}^{a}$, defined to be $A_{\mu \parallel}^{a}(x,z) \equiv
\partial_{\mu} \phi^{a}(x,z)$, at $q^2=m_{P}^{2}$ with boundary
conditions $ \phi^{a \prime} (z_{IR})=\pi^{a \prime}
(z_{IR})=\phi^{a}(0)=\pi^{a}(0)=0$ (no summation)
\begin{eqnarray}
&&\left( \partial_{z}^{2} + \partial_{z}\left(\ln a(z) \right) \partial_z \right) \phi^a +g_{5}^{2} a^2 (z) \left( M_{A}^{2} \right)_{aa} \left(\pi^a - \phi^a \right) =0, \nonumber\\
&&\partial_{z} \left( a^3(z) \left( M_{V}^{2} +M_{A}^{2} \right)_{aa} \partial_z \pi^a \right) + a^3(z) q^2 \left( \left( M_{V}^{2} +M_{A}^{2} \right)_{aa} \left(\pi^a- {1\over 2} \phi^a \right) + \left( M_{V}^{2} -M_{A}^{2} \right)_{aa} {1 \over 2} \phi^a \right) =0.\nonumber\\
\end{eqnarray}

The decay constants of massless pseudoscalar mesons $P^a$ (which
is a good approximation for pions) are given by
\cite{hep-ph/0501128}
\begin{equation}
f_{P^a}= - {1 \over g_{5}^2} {\partial_{z} A^{a}(0,z) \over z}
\arrowvert_{z=\epsilon},
\end{equation}
with $A^{a}(0,z)$ the solution of Eq.(\ref{eq.phiA}) satisfying
$A^{a \prime}(0,z_{IR})=0$ and $ A^a(0,\epsilon)=1$.

The results are shown in Table \ref{tab.axialresults} and Table \ref{tab.vectorresults}. It is well-known that
without flavour symmetry, the $1 {}^3P_{1}$ isodoublet state mixes
with $1 {}^1P_{1}$ isodoublet state. The mixing angle $\theta$ is
defined by parametrizing the physical state $K_{1}(1273)$ and
$K_{1}(1402)$
\begin{eqnarray}
\arrowvert K_{1}(1402) \rangle = \arrowvert K_{1A} \rangle \cos\theta -\arrowvert K_{1B} \rangle \sin\theta, \nonumber\\
\arrowvert K_{1}(1273) \rangle = \arrowvert K_{1A} \rangle
\sin\theta -\arrowvert K_{1B} \rangle \cos\theta,
\end{eqnarray}
such that
\begin{equation}
\sin^2 \theta =\left( {M^{2}_{K_{1}(1402)} -M^{2}_{K_{1A}} \over M^{2}_{K_{1}(1402)}-M^{2}_{K_{1}(1273)} } \right).\nonumber
\end{equation}
In the numerical analysis we fit the value of the quark mass and condensate to the case of ideal mixing ($\theta=45^o$).

We now turn to the case where the warp factor is deformed from the
$AdS$ case. We consider the case that only $\alpha_1$ is non-zero.
In this case the solution $\langle X(z) \rangle$ is of the form
$\langle X(z) \rangle  = \mathit{diag} (X_{u}(z), X_{u}(z),
X_{s}(z))$ where $X_{u}(z)$ and $X_{s}(z)$ approach their $AdS$
limit respectively when $z \to 0$. All the qualitative discussions
above still hold and our numerical results are shown in Table 3
and Table 4. $\alpha_1$ is expected to be of order $O( g_{5}^2
m_{q}^{2})$ from matching the two-point function to the same
function in the large $N_c$ QCD. Our numerical result for
$\alpha_1$ agrees with this.

\section{Numerical Analysis}

Precisely as in \cite{hep-ph/0501128} the value of the IR cutoff
is calculated using the equation of motion of the $\rho$ meson,
which is independent of the quark mass. The value of this cutoff
is however a function of the deformation parameter, $\alpha_1$,
which is tuned such that the $\rho$ decay constant is closer to
the experimental value. Note that by tuning $\alpha_1$, it is
possible to obtain exactly the correct value for this decay
constant. In this work a semi-global fit has been used so that the
value of the decay constant is good but not exact, such that the
predictions of other observables are closer to the experimental
values. This method is similar to the global fit used in
\cite{hep-ph/0501128} but here, because we have three more free
parameters, a full global fit is unmanageable. This presumably
means that a full survey of the parameter space would give even
better results than those presented here if such a fit were
computable.

Having calculated the IR cutoff, the excited states of the $\rho$
meson are found by studying the higher modes of the appropriate
five-dimensional field.

\begin{figure}[!h]
\begin{center}
\includegraphics[width=8cm,clip=true,keepaspectratio=true]{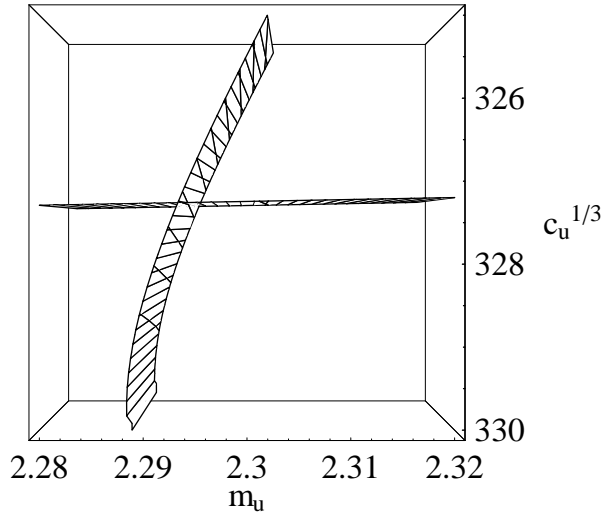}
\caption{Intersection of the physical contours for $m_\pi$ and $f_\pi$ in the $(m,c^{\frac{1}{3}})$ plane.}\label{fig.mcazero}
\end{center}
\end{figure}

In \cite{hep-ph/0501128}, a perturbative expansion is used to
prove the Gell-Mann-Oakes-Renner relation
\cite{Phys.Rev.175.2195}. We show here that an alternative
numerical approach can be used to calculate the values of the
quark mass $m$ and condensate $c$. The value of the pion decay
constant is a function of the quark mass and condensate. Similarly
the boundary behaviour of the five-dimensional field corresponding
to the pion is a function of these parameters (with $q^2$ set to
$m_\pi^2$ in the pseudoscalar equation of motion). We can plot
both of these quantities as a function of the quark mass and
condensate. Each plot should have a line in $(m,c)$ space
corresponding to the physical values (the correct value of $f_\pi$
and the correct boundary behaviour respectively). The point at
which these lines intersect therefore gives the correct value of
$m$ and $c$ to provide the physical pion mass and pion decay
constant. These ``physical contours" are shown in figure
\ref{fig.mcazero} to illustrate that this method produces almost
identical values to the perturbative approach
\cite{hep-ph/0501128}. In both the undeformed and deformed $AdS$
cases it can be shown that the GMOR relation holds to order
$m_q^2$ as expected.

The mass of the $a_1$ is calculated by solving the axial vector
equation of motion with the appropriate boundary conditions. It is
found that in the case of deformed $AdS$, the mass and decay
constant values for this axial vector particle are in remarkably
close agreement with experiment.

For the kaon sector, we perform a global fit in the space of $m_s$
and $c_s$ by studying the mass of the vector and axial vector
$K$-mesons corresponding to the $K^*(892)$ and $K_{1A}$ mesons. It
is possible to get within a few percent of the experimental value
(in the case of ideal mixing) though it may be possible by tuning
the deformation parameter and the values of $m$ and $c$ correctly
to get these masses exact.

This concludes the calculation of the free parameters of the
theory. Using these values, the decay constants for the isodoublet
and isosinglet sectors are calculated using Eq.(\ref{eq.VAdecay}).

We have shown by numerical analysis that even without a full
global fit, the vector sector is in close agreement with the
experimental results. Compared with the results of the QCD sum
rules, this holographic action appears to model the vector and
axial vector spectrum of the strong force to high precision. A
global fit would almost certainly reduce the average error to the
order of a percent or two, however as explained above such a
calculation would require considerable computing time.

We provide in Tables \ref{tab.parameters}, \ref{tab.axialresults}
and \ref{tab.vectorresults} the values calculated in our analysis.

\begin{table}[!h]
\centering
\begin{tabular}{||l|l|l||}
  \hline
  Parameter & Value (MeV) with $\alpha_1=225^2$ &Value (MeV) with $\alpha_1=0$\\
  \hline
  \hline
  $Z_{IR}^{-1}$ & 335.6 & 322.6 \\
  $m_u$ & 2.012  & 2.29 \\
  $c_u^\frac{1}{3}$ & $346.25$ & $327.25$\\
  $m_s$ & 90 & 100 \\
  $c_s^\frac{1}{3}$ & $457.53$  & $232.5$ \\
  \hline
\end{tabular}
\caption{Values chosen for the free parameters for both pure
$AdS_5$ and the deformed geometry. Note that in the case of
$\alpha_1\ne 0$, we have performed a semi-global fit over pairs of
variables.}\label{tab.parameters}
\end{table}

\begin{table}[!h]
\centering
\begin{tabular}{||l|l|l||}
  \hline
  Observables & Value (MeV) with $\alpha_1=225^2$ (\% error) &Value (MeV) with $\alpha_1=0$ (\% error)\\
  \hline
  \hline
  $m_\pi$ & $139.6*$& $139.6*$\\
  $f_\pi$ & $92.4*$ & $92.4*$\\
%  $m_\pi-\sqrt{\frac{2m_uc_u}{f_\pi^2}}$ & $-0.2755$  & 0* \\
  $m_{a_1}$ & 1238 (0.24) & 1363 (10.4)\\
  $\sqrt{F_{a_1}}$ & 434.92 (0.44) & 486 (12.2)\\
  $m_{K_{1A}}$ & 1339* & 1339*\\
  $\sqrt{F_{K_{1A}}}$ & 436 $(\sqrt{F_{K_1(1400)}}\sim 454^{\dagger})$ & 450 $(\sqrt{F_{K_1(1400)}}\sim 454^{\dagger})$\\
  $m_{A_3}$ & 1389  & 1339 \\
  $\sqrt{F_{A_3}}$ & 430 & 443 \\
  \hline
\end{tabular}
\caption{Axial sector results for both pure $AdS_5$ and the
deformed geometry using the parameters in table
\ref{tab.parameters}. Note that the $\alpha_1=0$ results have been
calculated using the same numerical techniques as in
\cite{hep-ph/0501128}. The case of ideal mixing ($\theta=45^o$)
has been chosen to fix the strange quark mass and condensate
values. Experimental values are chosen as the midpoint of those in
\cite{Phys.Lett.B592.1}. The decay constant of the $a_1$ is
compared with the lattice result \cite{Phys.Rev.D39.1357}. $*$
indicates that this value is used to fix the free parameters, all
other values are predictions. $\dagger$ results are taken from
\cite{Phys.Rev.Lett.74.4596}. The axial vector meson $A_3$
corresponds to the isosinglet meson in the
octet.}\label{tab.axialresults}
\end{table}

\begin{table}[!h]
\centering
\begin{tabular}{||l|l|l||}
  \hline
  Observables & Value (MeV) with $\alpha_1=225^2$ (\% error) &Value (MeV) with $\alpha_1=0$ (\% error)\\
  \hline
  \hline
  $m_\rho$ & 775.8* & 775.8*\\
  $\sqrt{F_\rho}$ & 335.5 (2.8) & 329 (4.5)\\
  $m_{\rho'}$ & 1830 & 1781\\
  $\sqrt{F_{\rho'}}$ & 635.38 & 616.4 \\
  $m_{K^*}$ & 793 (11)* & 799 (10.4)* \\
  $\sqrt{F_{K^*}}$ & 330.8 $(11^\dagger)$ & 329 $(11^\dagger)$ \\
  $m_{V_3}$ & $m_\rho$  & $m_\rho$\\
  $\sqrt{F_{V_3}}$ & $\sqrt{F_\rho}$ & $\sqrt{F_\rho}$\\
  \hline
\end{tabular}
\caption{Vector sector results for both pure $AdS_5$ and the
deformed geometry. The $\rho'$ is the first excited state of the
$\rho$ meson. Experimental values are chosen as the midpoint of
those in \cite{Phys.Lett.B592.1}.  $*$ indicates that this value
is used to fix the free parameters, all other values are
predictions. $\dagger$ results are taken from lattice predictions
\cite{hep-lat/9611021} though these have large uncertainties. The
vector meson $V_3$ corresponds to the isosinglet meson in the
octet.}\label{tab.vectorresults}
\end{table}
\newpage
\section{Discussion and Conclusion}
We have extended the work of Erlich et al \cite{hep-ph/0501128} to
include an $SU(3)$ flavour group with a distinct strange quark
mass. We also study the effects of a deformation of the metric by
the simplest possible factor. This deformation is sufficient to
provide a non-conformal region in the effective four-dimensional
theory. We show that the effect of adding this extra parameter is
enough to give very accurate ($\epsilon_{rms}<2\%$) masses and
decay constants for the light quark sector.

We extend the formalism with the addition of a strange quark and
calculate the vector and axial vector masses and decay constants
in the $s$ quark sector. The values of the strange quark mass and
condensate are tuned to give the $K_{1A}$ and $K^*$ masses to
within a few percent of the experimental values (though we assume
ideal mixing). This fit is a function of the deformation
parameter, $\alpha_1$, so that with a full global fit it may be
possible to lower the average error, which in this analysis is
$\sim 7\%$. With the mass and condensate values fixed, we
calculate the decay constants which are compared with lattice data
providing a good agreement.

Though in the sectors described in this paper an excellent
agreement with experiment and lattice is obtained, it is found
that the strange pseudoscalar (the kaon) gives a surprisingly poor
fit for the parameter values chosen here. It is also possible to
tune $m_s$ and $c_s$ such that the kaon mass is correct giving a
good fit for the $K^*$ but a poor result for the $K_1$.

Note that the values of $m_u$ and $c$ have been fixed without
reference to the kaon sector. It may well be possible to tune
these values such that the kaon sector provides a better fit to
the data. In this letter we merely wish to illustrate that these
extensions to the work of Erlich et al \cite{hep-ph/0501128} can
provide reasonable fits in the strange quark sector. We leave a
full global fit for further calculation.

There are several possible extensions to this work. It would be
interesting to study the mass splitting in the isovector spectrum
by the addition of different mass $u$ and $d$ quarks and a
four-dimensional $U(1)$ gauge interaction. It would also be
interesting to study the addition of an operator corresponding to
baryons in these models. Several other interesting directions are
noted in \cite{hep-ph/0501128} which should guide us even closer
to a true holographic dual of QCD.

\section{Acknowledgements}
Both Jonathan Shock and Feng Wu would like to thank the Project of
Knowledge Innovation Program (PKIP) of the Chinese Academy of
Science (CAS) for funding this work.

\end{document}